\documentclass[showpacs,preprintnumbers,amsmath,amssymb,12pt]{revtex4}
\usepackage{graphicx}
\input epsf
\usepackage{dcolumn}
\usepackage{bm}
\usepackage[pdftex,
            pdfauthor={Artyom V. Yurov, Valerian A. Yurov, Artyom V. Astashenok},
            pdftitle={The Cosmological Models with Jump Discontinuities},
            pdfsubject={},
            pdfkeywords={Classical cosmology, jump discontinuity, cosmological singularities},
            pdfproducer={Latex with hyperref},
            pdfcreator={pdflatex}]{hyperref}
\hypersetup{colorlinks=true, citecolor=blue, linkcolor=red}
\usepackage{enumerate}
\usepackage{scrextend}
\addtokomafont{labelinglabel}{\sffamily}
\newcommand{\beq}{\begin{equation}}
\newcommand{\beqn}{\begin{equation*}}
\newcommand{\enq}{\end{equation}}
\newcommand{\enqn}{\end{equation*}}

\newcommand{\R}{{\mathbb R}}

\renewcommand{\l}{\lambda}

\newtheorem{remark}{\textsc{Remark}}

\begin{document}
\newcommand{\be}{\begin{equation}}
\newcommand{\ee}{\end{equation}}
\newcommand{\bea}{\begin{eqnarray}}
\newcommand{\eea}{\end{eqnarray}}
\newcommand{\nn}{\nonumber \\}
\newcommand{\e}{\mathrm{e}}

\title{The Cosmological Models with Jump Discontinuities}
\author{Artyom V. Yurov}
\email{artyom\_yurov@mail.ru}
\author{Artyom V. Astashenok}
\email{artyom.art@gmail.com}
%
\address{I. Kant Baltic Federal University, Department of Physics, Mathematics and IT, Al. Nevsky str. 14, Kaliningrad
236041, Russia}
\author{Valerian A. Yurov}
\email{vayt37@gmail.com}
\affiliation{I. Kant Baltic Federal University, Functionalized Magnetic Materials for Biomedicine and Nanotechnology Center, Department of Physics, Mathematics and IT, Al. Nevsky St. 14, Kaliningrad, 236041, Russia}

\begin{abstract}
The article is dedicated to one of the most undeservedly overlooked properties of the cosmological models: the behaviour at, near and due to a jump discontinuity. It is most interesting that while the usual considerations of the cosmological dynamics deals heavily in the singularities produced by the discontinuities of the second kind (a.k.a. the essential discontinuities) of one (or more) of the physical parameters, almost no research exists to date that would turn to their natural extension/counterpart: the singularities induced by the discontinuities of the first kind (a.k.a. the jump discontinuities). It is this oversight that this article aims to amend. In fact, it demonstrates that the inclusion of such singularities allows one to produce a number of very interesting scenarios of cosmological
evolution. For example, it produces the cosmological models with a finite value of the equation of state parameter $w=p/\rho$ even when both the energy density and the pressure diverge, while at the same time keeping the scale factor finite. Such a dynamics is shown to be possible only when the scale factor experiences a finite jump at some moment of time. Furthermore, if it is the first derivative of the scale factor that experiences a jump, then a whole new and different type of a sudden future singularity appears. Finally, jump discontinuities suffered by either a second or third derivatives of a scale factor lead to cosmological
models experiencing a sudden dephantomization -- or avoiding the phantomization altogether. This implies that theoretically there should not be any obstacles for extending the cosmological evolution beyond the corresponding singularities; therefore, such singularities can be considered a sort of a cosmological phase transition.

\end{abstract}


\maketitle

\section{Introduction} \label{sec:Intro}

There are very few ideas that are as ingrained in the contemporary cosmology as that of a cosmological singularity. Although initially quite vehemently opposed by the very father of the General Relativity, Albert Einstein, by the end of the 1990-s this concept became so widespread and recognizable that it even gave its name to the most popular cosmological model of the day -- the Big Bang theory has derived its moniker from the so-called Big Bang initial singularity (BBS), out of which the entire expanding universe must have emerged in all its high-energy glory. The similar model of a collapsing universe was said to be evolving towards yet another singularity, called the Big Crunch singularity (BCS).

Both of these singularities occur when the scale factor $a(t)\to 0$ at $t\to t_s$ (the value $t_s$ will from now on denote the precise moment when a singularity arises). Prior to 1990-s most cosmologists were confident that the density and pressure of the kind of matter that fills the universe have to abide by either the strong or, at a worst case, the weak energy conditions. These conditions, filtered through the solutions of the Friedman-Lema\^{i}tre-Robertson-Walker equations (FLRW), led the cosmologists to believe that in all but two physically plausible scenarios the matter density and the pressure must diverge at the singularity: $\rho\to\rho_s=\infty$, $p\to p_s=\infty$; and so should all the derivatives of a scale factor: $|d^n a/dt^n|_{t\to t_s}\to\infty$ for
$n\ge 1$. The only two exceptions were the models of Milne (an open universe with $\rho, \lambda \equiv 0$) and of De Sitter (a flat universe with $\rho \equiv 0$ and a positive cosmological constant $\lambda$), but since both of them had a misfortune to be literally {\em empty} -- having zero energy density $\rho$ -- both of them were deemed to be not particularly interesting (their incompatibility with the observational data didn't help matter either).

The discovery of a cosmic acceleration  \cite{1} in 1998 changed all that. Suddenly there was an unequivocal observational proof that there is more in the Universe than the ordinary matter with its usual energy constrains. The universe was evidently filled to a brim with a weird new field, called a ``dark energy'', and all bets were off.  The proverbial floodgates were open. The cosmologists rolled up their sleeves and started rolling out one after another exciting new models aimed to describe and understand the universe filled with that or another version of a dark energy. And while some of those models did have the old familiar faces of BBS and BCS singularities, pretty soon it became obvious that these two are but a tip of a singular iceberg of possibilities. In particular, the cosmologists realized that the singularity should not be solely associated with zero values of a scale factor $a$, but in fact can occur when $a$ reaches infinity in finite time. And then they went even further and have found out the singular solutions with a {\em finite non-zero} scale-factor $0<a<\infty$! Similar to the BBS and BCS, each of the new singularities has received a befitting name: a Big Rip singularity (BRS) \cite{BR}, \cite{SF-3}, a Big Freeze singularity (BFS) \cite{BF}, \cite{Sahni}, a Sudden Future singularity (SFS) \cite{SF-1}, \cite{SF-2}, \cite{SF-3} a Big Boost singularity (BBtS) \cite{boost}, a Big Break singularity (BBS) \cite{BB}, \cite{BB1}, the $w$-singularities \cite{w-sing}, \cite{w-Fer}, the inaccessible singularities \cite{inas}, and the directional singularities \cite{direct}. A common feature in all of these models except for the singularity of type IV (according to the Nojiri-Odintsov-Tsujikawa classification of \cite{Odin-clas}) is that they all predict cosmic evolution to end by means of a curvature singularity, $|{\ddot a}(t)|\to\infty$, reachable in a finite proper time, say as $t\to t_s$.

Now, what are the requirements and characteristic properties of all these solutions?.. The BRS takes place in the phantom models with $a\to\infty$ for the
$t\to t_s$, where $a=a(t)$ is the scale factor. The models with SFS, BFS, BBtS and BBS all predict a finite value of the scale factor $a_s=a(t_s)<\infty$ but they differ in their own distinct values of the Hubble expansion parameter $H_s=H(t_s)$ as well as in different signs of (divergent) expression ${\ddot a_s}/a_s$ (see also
\cite{Oddd}).  Yet another type of singularity, aptly called the $w$-singularity, was obtained in \cite{w-sing} and it is characterized by the finite scale factor, vanishing pressure and energy density (similar to the the type IV cosmological singularity), so that the only thing singular is a time-dependent barotropic
index $w(t)$. It is important to note that the $w$-singularities differ from the singularity of IV type in that it does not exhibit the divergence of the higher derivatives of the Hubble parameter (\cite{w-sing}).

At this point one can ask: is it possible to name a property that would unite all these distinct models?.. Surprisingly, the answer is yes. These singularities all share one similar trait: each and every one of them is associated with an {\em essential discontinuity}. It is this feature that should give a pause to an attentive reader who at this point may ask: but aren't the essential discontinuities just {\em one} of the type of possible discontinuities?.. Or, to put it in other words: is it not reasonable to assume that in our fixation on one special type of discontinuities we might actually be missing some other interesting phenomena?.. Just to illustrate our point, consider a singularity of IV type by N.O.T. classification \cite{Odin-clas}. Such singularities are associated with essential discontinuity of the third order derivative of the scale factor. But what about the derivatives of the lower order?.. One usually assumes that the second derivative here is properly defined and continuous, even at the moment of singularity itself. But it is not the only possibility. The divergence of the third derivative at $t=t_s$ might as well imply that the second derivative has a jump discontinuity at $t\to t_s$!.. Similar reasoning can also be provided for other types of singularities. So, it is the goal of this article to correct this glaring oversight by examining these exotic cases. Admittedly, at first glance this line of research might not look promising. We can even go as far as to predict two typical objections issued by an imaginary referee:

{\bf Objection 1.} If n-th derivative of the scale factor diverges  $a^{(n)}(t_s)=\pm\infty$ and $|a^{(n-1)}(t_s)|<\infty$ in the moment of singularity,
the existence of a jump discontinuity for $a^{(n-1)}(t)$ might be completely inessential for the dynamics of the universe and hence lead to no new observable phenomena.

{\bf Objection 2.} The existence of a jump discontinuity might seem natural from the purely mathematical
point of view, but what possible {\bf physical} meaning can lie beyond such a wild assumption?

Naturally, the list of the arguments can go on, but it is these two arguments we consider most pertinent and most deserving of an answer.

In this article we will concentrated on dispelling the first of these objections, for an obvious reason. If the addition of a jump discontinuity to our solutions would produce no new results, such an addition should be deemed completely superfluous -- and there would really be no point at all in discussing the second objection. Fortunately, that is not our case. As we shall see the solutions with the jump discontinuities {\em do} have new properties, making them distinct from the known solutions that only have the essential discontinuities.

Still, the second objection remains valid. That any physical field (say, a scalar field $\phi$ with a potential $V(\phi)$ governed by an ordinary power law) might produce a cosmological dynamics characterized by a jump discontinuity -- doesn't it seem to be completely impossible?.. Well, the answer is no, and we can prove it with a simple counterexample.

Consider the aforementioned model of a flat Friedmann universe containing a scalar field $\phi=\phi(t)$ and a common quadratic potential $m \phi^2/2$. In addition, let us also assume that the universe in question is filled with a {\em negative} cosmological constant, finely tuned with the ``mass'' $m$ of a scalar field in a following very specific (but not unreasonable!) way:

\beq \label{V}
V(\phi) = \frac{9 \lambda^2}{4} (\phi-\phi_0)^2 - \frac{\lambda^2}{2}.
\enq

The cosmological dynamics of such a universe will follow the standard Friedmann equations:
\beq \label{Fried-V}
\begin{split}
\ddot \phi &= - 3 H \dot \phi  -\frac{d V(\phi)}{d \phi},\\
H^2 &= \frac{1}{2}\dot \phi^2 + V(\phi), \\
\dot H &= - \frac{3}{2} \dot \phi^2.
\end{split}
\enq
The system \eqref{Fried-V} with the potential \eqref{V} can be solved via the reduction to the Abel equation \cite{YY10}; the resultant solution will have the following form:
\beqn
\begin{split}
\phi &= \phi_0 + \lambda |t| = \phi_0 +  \kappa \lambda t, \\
H &= \frac{3}{2}\kappa \l (\phi_0 - \phi) = -\frac{3 \l^2}{2} t,
\end{split}
\enqn
where $\kappa = +1$ for $t>0$, and $\kappa=-1$ for $t<0$. Note here that while both the Hubble parameter $H$ and the scalar field $\phi$ are continuous, the first derivative of the latter experiences a jump discontinuity at $t=0$! This vividly demonstrates that even the most respectable potentials $V(\phi)$ might, under the right circumstances, begin to behave themselves in a way consistent with our hypothesis and to produce some sort of jump discontinuities.

In this work we will demonstrate four important features the cosmological solutions with jump discontinuities might possess:

\begin{itemize}
\item It is possible to construct the cosmological solutions with $w_s=p_s/\rho_s<\infty$ although $p_s=\infty$, $\rho_s=\infty$,
$a_s<\infty$;

\item The energy density can diverge ($\rho_s=\infty$) while the pressure remains finite ($p_s<\infty$). In other words, the Lagrangian for
such models doesn't diverge at the moment of singularity  (Sec. \ref{sec:w_c});

\item New cosmological solutions with SFS appear (Sec. \ref{section:SSS});

\item Cosmological solutions with jump discontinuities might avoid an effect of phantomization (Sec. \ref{sec:dephantom}).
\end{itemize}

Finally, in Sec. \ref{sec:conclusion} we will briefly discuss the possible importance of the jump points in the cosmology and will argue that their existence has to be accepted, if only to avert much more serious problems with a behaviour of physical fields when the universe enters the phantom zone. \footnote{One should also note that it is not the first time the jump discontinuities enters the cosmological consideration: such discontinuities are a must for all the brane models (cf. \cite{YY}).}

We will subsequently study three types of cosmological models:

\begin{labeling}{Model (iii)}
\item [Model (i)] whose scale factor $a$ has a singular point of the first type (i.e. a jump discontinuity) at $t=t_s$;

\item [Model (ii)] in which the first derivative of a scale factor has such a point at $t=t_{s}$ and

\item [Model (iii)] in which the second (or higher) order derivative of a scale factor has a jump at $t=t_{s}$.
\end{labeling}

Let us see now what these models look like and what properties they possess.

\section{Cosmological models with $w^c$-singularities} \label{sec:w_c}

We begin by considering a spatially flat universe with the FLRW metric
\be
ds^2=dt^2-a(t)^2\left (dx^2+dy^2+dz^2\right).
\ee
The Friedmann equations for such a universe have the form\footnote{Throughout the article we will be using the ``natural'' system of units in which $8\pi G/3 = c = 1$}
\be
\frac{\dot{a}^{2}}{a^{2}}=\rho,\quad
\frac{\ddot{a}}{a}=-\frac{1}{2}\left(\rho+3p\right),
\ee
where $\rho$ and $p$ are the total energy density and the pressure.

As we have discussed in the end of Sec. \ref{sec:Intro}, we will study three possibilities. In case (i) (the scale factor itself experiences the jump) the density $\rho$ must necessarily diverge: $\rho\to+\infty$. This necessity, however, does not extend on the behaviour of pressure $p$: it might either stay finite $p\rightarrow p_s$ or it can diverge $|p|\rightarrow\infty$ at $t=t_{s}$. These properties closely mimic those of a singularity of the III type according to the N.O.T. classification. The type III usually arises in models with the equation of state \cite{Stefancic}
\begin{equation}
p=-\rho-A\rho^{k}. \label{Stef}
\end{equation}
But our model is quite different, because in it:
\begin{enumerate}
\item All energy conditions (weak, strong and dominant) can actually hold at $t=t_s$;

\item The barotropic index $w=w(t)=p(t)/\rho(t)\to \infty/\infty = w_s={\rm const}<\infty$ or $w\rightarrow 0$ (the latter holds when, for example, pressure $p$ remains finite as $t \to t_{s}$). In particular, it is possible to obtain $w_s=0$ even in a case when pressure diverges.
\end{enumerate}

It is useful to think about the second property (when $|p|\rightarrow\infty$ at $t\rightarrow t_{s}$) in view of a duality existing between a Big Bang and a $w$-singularity \cite{w-sing}. Namely, the BBS is characterized by $p\to\infty$, $\rho\to\infty$ and $w\to 0$ for the universe filled with a baryon and/or dark matter whereas the $w$-singularity is characterized by $p\to 0$, $\rho\to 0$ and $w\to \infty$. In contrast to the Big Bang (and the Big Crunch) singularity where $a\to 0$ when $t\to t_s$, in our model $a\to a_s$, $0<a_s<\infty$. Since the term ``duality'' has already been used for the purpose of comparison between the $w$-singularity with BBS, we will instead use the term ``conjugate'' to define our new type of singularity, and will henceforth call it a {\em conjugate $w$-singularity} or $w^{c}$-singularity.

\begin{remark}
We'd like to stress that from a physical point of view the $w$-singularity in not a proper singularity because all the observable functions (density,
pressure and the higher derivatives of the scale factor or the Hubble parameter $H$) stay finite even at $t=t_s$. Moreover, the very definition of $w$-singularity
given in \cite{w-sing} is actually incomplete. To show this lets consider the following form of the scale factor
\begin{equation}
a(t)=a_s-A\left(t_s-t\right)^m, \label{w-contr}
\end{equation}
where (\ref{w-contr}) is the special case of the general form of the scale factor from the \cite{w-sing} (with $B=0$, $A=a_s$,
$C/t_s^n=-A$, $D=1$, $n=m$). One can show that for the $t\to t_s$ we get
\begin{enumerate}
\item Type III singularity if $0<m<1$;

\item Type II singularity if $1<m<2$;

\item $w$-singularity if $m>2$.
\end{enumerate}

Furthermore, we have two special cases: $m=1$ and $m=2$. The case $m=1$ corresponds to a model with a constant barotropic index $w=-1/3$. The case $m=2$ is the most interesting one because
$$
\rho\to 0,\qquad p\to\frac{4A}{3a_s}\ne 0,\qquad |w|\to\infty.
$$
and
$$ \frac{d^{2n}H}{dt^{2n}}=0,\qquad
\frac{d^{2n+1}H}{dt^{2n+1}}\sim \frac{A^{n+1}}{a_s^{n+1}}<\infty,
$$
at $t=t_s$. Thus we end up with a sort of a generalization of $w$-singularity with a finite non-vanishing pressure at $t=t_s$.

In contrast to $w$-singularity, $w^c$-one is a true cosmological singularity since density (and, depending on the model, pressure) diverges at $t=t_s$.
\end{remark}

It is easy to see that $w^{c}$-singularity is irreproducible in the models with a continuous scale factor $a(t)$. To demonstrate this, we should start by constructing a  singularity of the III that corresponds to equation of state (\ref{Stef}). To satisfy the weak energy condition we choose $A<0$ and $\rho>0$. Thus the barotropic index
has the form
\begin{equation}
w(t)=-1+|A|\rho^{k-1}. \label{w-stef}
\end{equation}
Since for the singularity of the III type $\rho\to\infty$ and for the $w^{c}$-singularity $w\to w_s<\infty$ it is necessary to pick $k<1$ (in
the case $k=1$ we have a standard cosmology with a constant barotropic index (\ref{w-stef}) which does not result in a singularity with a finite value of the scale factor). After one integration one gets
$$
\log\frac{a(t)}{a_0}=-\frac{\rho^{1-k}}{3(1-k)|A|},
$$
so $a(t)\to 0$ when $\rho(t)\to\infty$. Thus in framework of the model \cite{Stefancic} one gets a Big Bang/Crunch singularity rather than $w^{c}$-one.

This conclusion will also hold for a general case of an analytic scale factor. Indeed, if $a(t_s)=a_s$ and $a(t)$ is a continuous function then for a sufficiently small open neighbourhood of $t_s$
$$
a(t)\to
a_s+\mu\left(t_s - t\right)^{\nu}+o\left(|t_s-t|^{\beta}\right),
$$
with $\beta>\nu>0$. Therefore the barotropic index will have a form
$$
w(t)\to\frac{2a_s(1-\nu)}{3\mu\nu(t_s-t)^{\nu}}\to\infty,
$$
and a chance to have a $w^{c}$-singularity once again slips right through our fingers.

For further description of the models with $w^{c}$-singularities it will be useful to define a {\em jump function}:

\begin{equation} \label{fried}
U(t)=
\begin{cases}
0,& {\rm
if} ~ t<t_s
\\
A/2, & {\rm if} ~ t=t_s
\\
A, & {\rm if} ~ t>t_s
\end{cases}
\end{equation}
There are few analytic representation of (\ref{fried}), for
example
\begin{equation}
\qquad \qquad \qquad \qquad U_1(t)=\frac{A}{2}\lim_{\alpha\to
+\infty}\left(1-\frac{2}{\pi}\arctan\left(\alpha(t_s-t)\right)\right),
\label{U1}
\end{equation}
\begin{equation}
\displaystyle{ U_2(t)=\lim_{\alpha\to +\infty}\frac{A}{1+{\rm
e}^{\alpha(t_s-t)}}}, \label{U2}
\end{equation}
\begin{equation}
\displaystyle{ U_3(t)=A\lim_{\alpha\to +\infty} 2^{-{\rm
e}^{\alpha(t_s-t)}}}. \label{U5}
\end{equation}
For the $U_1$, $U_2$, $U_3$ one gets
\begin{equation} \label{fried-1}
{\dot U_{i}}(t)=
\begin{cases}
0, & {\rm if} ~t\ne t_s,
\\
{\rm sgn}\,A\times\infty
 & {\rm if} ~t=t_s.
\end{cases}
\end{equation}

If we are to differentiate (\ref{U1}), (\ref{U2}), (\ref{U5}) once more, we will immediately conclude that the second
derivatives of $U_1(t)$ and $U_2(t)$ are zero for any values of
$t$ including $t=t_s$. For the $U_3(t)$ the situation is a bit more complicated:
\begin{equation} \label{fried-2}
{\ddot U_{3}}(t)=
\begin{cases}
0, & {\rm if} ~t\ne t_s,
\\
-{\rm sgn}\,A\times\infty
 & {\rm if} ~t=t_s.
\end{cases}
\end{equation}

Let $\psi=a^n$, $n>0$ so the density and the barotropic index are
\begin{equation}\label{rhow}
\rho=\frac{\dot\psi^2}{n^2\psi^2},\qquad
w=\frac{p}{\rho}=-1+\frac{2n}{3}\left(1-\frac{{\ddot\psi}\psi}{\dot\psi^2}\right).
\end{equation}

Let ${\tilde a}(t)$ be a smooth function that describes the evolution of the observable universe. For example, one can choose
\begin{equation}\label{shh}
{\tilde
a}(t)=a_0\left[\frac{1-\Omega_{_\Lambda}}{\Omega_{_\Lambda}}\sinh^2\left(\frac{t}{T}\right)\right]^{1/3},
\end{equation}
where $a_0={\tilde a}(t_0)$ is the present value of the scale factor, $t_0$ -- the present time (i.e. time passed since the initial Big Bang singularity at $t=0$) and
$$
t_0=\frac{2\,{\rm
arctanh}\sqrt{\Omega_{_\Lambda}}}{3H_{_\Lambda}},\qquad
T=\frac{2}{3H_{_\Lambda}},\qquad
H_{_\Lambda}=\sqrt{\Omega_{_\Lambda}}H_0,
$$
where $\Omega_{_\Lambda}=0.72\pm 0.04$ and $H_0=72\pm 8$ km/s Mpc$^{-1}$ is the present value of the Hubble constant. The expression \eqref{shh} describes the universe filled with both dark matter (with density $\rho_{_M}=(1-\Omega_{_\Lambda})\rho_{crit}$) and a vacuum energy.

As a next step, lets define $\psi=\psi(t)=f(t)+U(t)$, where $f(t)={\tilde a}^n(t)$ and $U(t)$ is one of the functions from the set (\ref{U1}),
(\ref{U2}), (\ref{U5}). Since all these functions together with their derivatives are equal to zero at $t<t_s$, we can safely conclude that the dynamics
of a universe with the scale factor $a(t)=\psi^{1/n}(t)$ will be exactly the same as the dynamics of a universe with the scale factor
${\tilde a}(t)$ at $t<t_s$. Moreover, since all quantum corrections depend on the higher derivatives of the scale factor (for example, if the conformal fields  are dominant, one has to take into account the third derivative of the Hubble parameter), the quantum dynamics will be the same up to the moment $t=t_s$ where the term $U(t)$ will be dominant. Calculating the derivatives and substituting them into the (\ref{rhow}) one gets at
$t=t_s$: $\rho_s=\infty$, $|p_s|=\infty$ but
$w_s=p_s/\rho_s<\infty$, which of course means that we have a $w^c$-singularity. At exactly the moment of singularity the scale factor will be
$$
a_s=\left[\left({\tilde
a}(t_{s})\right)^n+\frac{A}{2}\right]^{1/n}<\infty,
$$
while the barotropic index will either be
\begin{equation}
w_s=-1+\frac{2 n}{3}, \label{w1}
\end{equation}
if $U$ if defined as (\ref{U1}) and (\ref{U2}), or
\begin{equation}
w_s=-1+\frac{2n(f(t_s)+A/2-f(t_s)\ln 2+A\ln 2/2)}{3A\ln 2},
\label{w2}
\end{equation}
if it is defined by (\ref{U5}).

One can choose the parameter $n$ (in case of (\ref{w1}) or $n$ and $A$ together in order to conserve all the energy conditions, excluding the dominant
one which is violated at the threshold of this singularity.

Below we consider two particular examples of cosmological models with $w^{c}$-singularities. The first one illustrates the case when both the density and the pressure diverge at the moment of singularity, whereas the second example describes $w^{c}$-singularity with a finite pressure.

{\bf Example 1.} First we introduce a little generalization of the function
(\ref{U2}):
\begin{equation}
\displaystyle{ U_4(\beta;t)=\lim_{\alpha\to
+\infty}\frac{A}{1+\beta{\rm e}^{\alpha(t_s-t)}}}. \label{U5b}
\end{equation}
Choosing $n=1$ and using (\ref{rhow}) one can calculate density, pressure and parameter of EOS $w$. For the equation-of-state parameter we have the following expression
\begin{equation}
w_{s}=-\frac{1}{3}-\frac{2}{3\beta}\frac{(\beta^{2}-1)(f(t_{s})+A/(1+\beta))}{A}
\end{equation}
Therefore the parameter $w_{s}$ is explicitly determined by free parameters $A$ and $\beta$. Choosing
$$
A=\frac{2f(t_s)(\beta^2-1)}{2-3\beta},
$$
one gets
$$
w_{s}=0.
$$
One might get a bit concerned by the fact that the pressure $p_{s}=\infty$:
\begin{equation}
p=\frac{4}{3}\frac{\beta-1}{\beta+1}\frac{\dot{f}(t_{s})}{f(t_s)}\frac{3\beta-2}{\beta}\alpha,\quad
\alpha\rightarrow\infty.
\end{equation}
However, the energy density $\rho\sim\alpha^{2}$ so it balances nicely with the pressure and we still have $w_{s}=0$. It is interesting to note that for any given $w=w_{0}$ we can choose the parameters $A$ and $\beta$ so that $w_{s}=w_{0}$. One interesting case occurs when
$$
w_{0}=-\frac{1}{3}-\frac{2}{3}\frac{\ddot{f}(t_{s})f(t_{s})}{\dot{f}^{2}(t_{s})}.
$$
Therefore for any giving $\beta$ we can choose the value of ``jump'' $A$ so that in the moment of singularity the equation of state parameter is continuous.
\newline

{\bf Example 2.} We have already discussed how the condition of continuity of $a(t)$ acts as a safeguard from appearance of a finite-time, future singularity with $p_s<\infty$, $\rho_s=\infty$. In order to get one, however, it suffices to assume that $a(t)$ has a discontinuity of the first type. This is interesting because the
finiteness of pressure implies the finiteness of the Lagrangian of matter.

Lets consider a following modification of (\ref{U5b}):
\begin{equation}
U_{5}(\beta,\gamma;t)=\lim_{\alpha\to
+\infty}\frac{g(t)}{1+\beta{\rm e}^{\alpha(t_s-t)}},
\end{equation}
where $\gamma$ is a constant and the function $g(t)$ is finite at point $t=t_{s}$. The requirement of finiteness of pressure at the moment
of singularity leads to the following conditions on function $g(t_{s})$ and on its first derivative:
\begin{equation}\label{gap}
g(t_{s})=\frac{2f(t_s)(\beta^2-1)}{2-3\beta},\quad
\frac{\dot{g}(t_{s})}{g(t_{s})}=\frac{2-3\beta}{2}\frac{\dot{f}(t_{s})}{f(t_{s})}.
\end{equation}

These conditions would be satisfied if
$$g(t)=Ae^{\gamma(t_{s}-t)}, \quad
A={2f(t_s)}\frac{\beta^2-1}{2-3\beta}, \quad
\gamma=\frac{3\beta-2}{2}\frac{\dot{f}(t_s)}{f(t_s)}.$$

The permissible set of values for parameter $\beta$ includes $2/3<\beta<1$ and $\beta<0$ (otherwise one would end up with a meaningless result for a scale
factor at the very moment of singularity, $a(t_{s})<0$).

In this case the value of pressure at the moment $t=t_{s}$ is finite:
\begin{equation}
p(t_{s})=\frac{2-3\beta}{3\beta}\left(\frac{2\ddot{f}(t_{s})}{f(t_{s})}+(3\beta-2)\frac{\dot{f}^{2}(t_{s})}{f^{2}(t_{s})}\right)
\end{equation}
Moreover, one can in principle choose $\beta$ in such a way that the pressure is continuous at the moment of singularity i.e.
$p(t_{s})=\tilde{p}(t_{s})$, where $\tilde{p}(t_{s})$ is the value of pressure for a solution $a(t)=\tilde{a}(t)$. For de Sitter solution
$\tilde{a}(t)\sim \exp(\sqrt{\Lambda}t)$ this would not be possible. But for (\ref{shh}) the condition of ``continuity'' for pressure at
$t=t_{s}$ reads as
$$
\tanh^{2}\left(\frac{t_{s}}{T}\right)=(1-\beta)(1+2\beta)
$$
From this relation for $t_{s}$ as a function of parameter $\beta$ it follows that
\begin{equation}
t_{s}=\frac{T}{2}\ln\left(\frac{1+x}{1-x}\right),\quad x=\left((1-\beta)(1+2\beta)\right)^{1/2}.
\end{equation}
The direct comparison of this relation with the conditions on the value of $\beta$ written above leads us to conclusion that $-1/2<\beta<0$ or $2/3<\beta<1$. For a cosmology with a scalar field we therefore would have a model with a Lagrangian that remains finite even at the very moment of singularity.

\section{Cosmological models with a survivable sudden future singularity (SSS)} \label{section:SSS}


Let's consider the case when the first derivative of a scale factor has a jump at the moment $t=t_{s}$.
\begin{equation}\label{DG}
\dot{a}(t_{s}+\epsilon)-\dot{a}(t_{s}-\epsilon)=\pm\alpha^{2}
a(t_{s}),\quad \epsilon\rightarrow 0.
\end{equation}
The question arises: what would happen with a scale factor at $t>t_{s}$? To answer this question one should first note that this case
corresponds to a sudden future singularity since the second derivative
of a scale factor diverges here: $\ddot{a}(t_s)\rightarrow\infty$. Therefore the
pressure $p$ at $t=t_{s}$ should also diverge while the energy density remains finite and unaffected.

In the case of (\ref{DG}) the solution for a scale factor for $t>t_{s}$ can be written as
\begin{equation}\label{CSS2}
a_{+}(t)=a(t_{s})g(t),
\end{equation}
The condition of continuity at $t=t_{s}$ gives $g(t_{s})=1$. The
condition (\ref{DG}) reads as
\begin{equation}\label{GAP-1}
\dot{g}(t_s)-\frac{\dot{a}_{-}(t_{s})}{a_{-}(t_s)}=\pm{\alpha^2}.
\end{equation}

Let's choose a solution for the scale factor in form (\ref{shh}). Then for every $t$ in a sufficiently small left neighbourhood of $t_s$ we'll have the following approximation
\begin{equation}\label{EXP}
a_{-}(t\approx t_{s})\approx
a_{s}-a_{0}\left(\frac{1-\Omega_\Lambda}{\Omega_\Lambda}\right)^{1/3}\frac{2}{3T}\sinh^{-1/3}\left(\frac{t_s}{T}\right)\cosh\left(\frac{t_s}{T}\right)(t_{s}-t).
\end{equation}
The subscript ``$-$'' is added to emphasize that this is a solution for the interval $0\leq t<t_{s}$. According to the classification of FLRW models presented in
\cite{Geodesics}, the linear expansion (\ref{EXP}) corresponds to a case of a weak singularity. We remind the reader that a singularity is called weak if its tidal forces aren't capable of disrupting a finite object falling into it. For a weak singularity the causal geodesics are complete and therefore the cosmological evolution in principle may be extended beyond the threshold of singularity -- hence the name ``survivable sudden singularity'' (SSS).

The condition (\ref{GAP-1}) for (\ref{DG}) reads as
\begin{equation}\label{GAP}
\dot{g}(t_{s})-\frac{2}{3T}\coth\frac{t_{s}}{T}=\pm\alpha^{2}.
\end{equation}
The simplest case occurs when the energy density remains continuous at $t=t_{s}$. The sign of $\dot{a}(t)$ changes but its absolute value
remains invariant. For example, one can write the function $g(t)$ as
\begin{equation}\label{CSS1}
g(t)=\frac{1}{\sinh^{2/3}\left(\frac{t_{s}}{T}\right)}\sinh^{2/3}\left(\frac{2t_{s}-t}{T}\right).
\end{equation}

Using the equation (\ref{CSS1}) with the condition (\ref{GAP}) produces the following equation on $T$ (and therefore on the value of the cosmological
constant):
\begin{equation}\label{EQL}
\frac{1}{T}\coth\left(\frac{t_{s}}{T}\right)=\frac{3}{4}\alpha^{2}.
\end{equation}
This equation has a single solution at $\alpha^{2}t_{s}\geq 4/3$ and no solutions at all if $\alpha^{2}t_{s}<4/3$. For $\lambda
t_{s}\rightarrow\infty$ the solution (\ref{EQL}) is $T=4\alpha^{2}/3$. The solution (\ref{CSS2}) describes a shrinking universe filled with the vacuum energy and cold dark matter. The cosmological constant is the same as was in the solution $a_{-}$.

One can note that the model we just considered actually coincides with the $\Lambda$CDM cosmology at $0\leq t< t_{s}$ and can therefore be used to describe and explain the data of the astronomical observations. However, there is an interesting twist here. While in a case of $\Lambda$CDM model the value of the cosmological constant can only be determined via the observations, in case of a model with SSS the said value can be determined right from the equations.

\begin{remark}
It is actually possible to construct many solutions that would satisfy the conditions (\ref{GAP}) and
$\rho_{+}(t_{s})=\rho_{-}(t_{s})$. For example, let's choose
\begin{equation}\label{CSS3}
g(t)=C(t_{f}-t)^{2/3}.
\end{equation}
The constants $C$ and $t_{f}$ can be derived from the conditions $g(t_{s})=1$, (\ref{GAP})
\begin{equation}
C=\left(3\alpha^2-(2/T)\coth(t_{s}/T)\right)^{2/3},\quad t_{f}=t_{s}+\left(3\alpha^{2}-(2/T)\coth(t_{s}/T)\right)^{-1}.
\end{equation}
The more rigid condition $\rho_{+}(t_{s})=\rho_{-}(t_{s})$ leads to the equation (\ref{EQL}), so for $C$ and $t_{f}$ we end with:
$$
C=\left(\frac{3}{2}\right)^{2/3}\alpha^{4/3},\quad t_{f}=t_{s}+\frac{2}{3}\alpha^{-2}.
$$
In contrast to (\ref{CSS2}) the solution (\ref{CSS3}) corresponds to a universe filled with a cold dark matter only. At the moment $t=t_{f}$ this universe ends its existence in the Big Crunch singularity.
\end{remark}

It is clear that one can choose a solution for $g$ in the form
\begin{equation}\label{CSS4}
g(t)=\frac{C}{a(t_{s})}\sinh^{2/3}\left(\frac{t_{f}-t}{T^{*}}\right)
\end{equation}
with constants $C$, $t_{f}$ and $T^{*}$. If we require continuity of energy density at $t=t_{s}$ these constants can be determined from the equations
\begin{equation}
C\sinh^{2/3}\left(\frac{t_{f}-t_{s}}{T^{*}}\right)=\sinh^{2/3}\left(\frac{t_{s}}{T}\right),
\end{equation}
\begin{equation}
\frac{C}{T^{*}}\coth\frac{t_{f}-t_{s}}{T^{*}}=\frac{3}{4}\alpha^{2}.
\end{equation}
For a given $C$ one can get the time before the singularity $t_{f}-t_{s}$ and the value $1/T^{*}$. The solution (\ref{CSS4})
describes the universe filled with the cold dark matter and the vacuum energy. But the value of the cosmological constant ends up being different from its original pre-singularity value (for $0<t<t_{s}$). Therefore, the following interpretation of this type of singularity is possible. At the moment $t=t_{s}$ the phase
transition occurs and a certain amount of the vacuum energy turns into the dark energy or vice versa. Although the possibility of such transformation is quite unclear from the physical point of view, there is also no clear physical law that might forbid a transition like that, so such a possibility should not be simply dismissed without at least a consideration.

\section{Cosmological models with a (de)phantomization and the jump discontinuities for the derivatives of a scale factor} \label{sec:dephantom}

In the previous section we have explored a model that can serve as a sort of a generalization of the classical $\Lambda$CDM model. In this section we will delve even deeper along the scale of the barotropic index straight to the weird domain of the phantom cosmologies with $w<-1$. Specifically, we will try to work out the answer to the following question: what might be the fate of the universe filled with phantom fields if at least one of the physical fields (say, $a$ or one of its time derivatives) would be allowed to have a jump discontinuity. As we shall see, this one seemingly minor assumption might have profound consequences for the entire evolution of the universe, allowing it to spontaneously ``dephantomize'', i.e. safely exit the so-called phantom zone.

Interestingly, the connection between the jump discontinuity of the physical fields and the problem of ``dephantomization'' has been known at least since the paper \cite{AYurov}. In it the model of a flat universe filled with a scalar field $\phi$ and the potential $V(\phi)$ was considered. It is known that in such a universe the density $\rho$ and pressure $p$ depend on the field $\phi$ as
\beqn
\rho = \frac{1}{2} \dot \phi^2 + V(\phi), \qquad p = \frac{1}{2} \dot \phi^2 - V(\phi),
\enqn
and the FLRW equations for the field $\phi$ and the Hubble parameter $H$ are
\beq \label{FRLW_phantom}
\begin{split}
\dot \phi^2 &= - \frac{2}{3} \dot H, \\
H^2 &= \rho.
\end{split}
\enq

From the first equation of \eqref{FRLW_phantom} it follows immediately that if $\dot H$ ever crosses zero and becomes positive, the kinetic term $\dot \phi^2$ becomes negative, effectively switching the range of $\dot \phi$ from the real to the imaginary numbers. However, the second equation from \eqref{FRLW_phantom} stalwartly guarantees that, as long as the absolute value of the kinetic term in $\rho$ does not exceed the value of potential $V(\phi) >0$, the physical field $H$ will stay real. Since the requirement $\dot H >0$ means that the universe undergoes a super-inflation, similar to the one produced by the phantom fields (i.e. the fields with the parameter of state $w<-1$), the described phenomena -- a change in the sign of the kinetic term while the rest of the physical fields stays real-valued -- is called ``phantomization''.

Naturally, the very real problem of a physical quantity suddenly becoming (at least temporarily) imaginary to day remains a highly controversial point of contention among the cosmologists. There were a number of interesting attempts to explain this over the years, in particular by extending the class of physically meaningful Hamiltonians to include the non-Hermitian ones \cite{ACK-1}, \cite{ACK-2}. However, as we shall see, there exist a different, less radical alternative: the universe might literally ``skip over'' the seemingly inevitable phantomization, similar to a frog jumping over the obstacle.

In order to demonstrate this, let's first assume that the time the universe spends in the ``phantom zone'' is finite (the assumption that is substantiated by the direct calculations for many physically significant types of potential $V(\phi)$). Then, as \cite{AYurov} argues, it is actually possible to completely {\em cut} the phantom region out of the model by taking not just one but two different solutions of \eqref{FRLW_phantom} $a_-(t)$ and $a_+(t)$ and {\em sewing} together at the points when $\dot H$ hits zero, thus producing a new scale factor $a$ which will be a proper solution of \eqref{FRLW_phantom} -- and in so doing we can actually prove that this $a$, its first and second derivatives, corresponding $\rho$ and its first derivative and even $p$ will all be continuous. This amazing feat is made possible by two important facts: first, that the general solution $a=a(t)$ of the equation \eqref{FRLW_phantom} always contains two constants of integration (recall that $H=\dot a / a$) and, second, that the points needed to be sewn together can be shown to be the inflection point of the function $\ln a$ (see \cite{AYurov} for more details).

For example, let's consider a particular model of the flat universe that for large $t \gg 0$ behaves as a classical FLRW universe filled with cold dust ($w=0$) and thus have the scale factor obey the power law $a(t)\approx t^{2/3}$. It is easy to see from \eqref{FRLW_phantom} that the simplest potential $V$ that leads to such a dynamics is:
\beqn
V(\phi(t)) = \frac{2}{9 t^2},
\enqn
and the corresponding general solutions of the \eqref{FRLW_phantom} for this potential are:
\beqn
a = \sqrt[3]{\frac{c_1}{t} + c_2 t^2}, \qquad p=-\frac{c_1\left(c_1+4 c_2t^3\right)}{3t^2\left(c_1+c_2t^3\right)^2},\qquad
\rho=\frac{\left(2 c_2 t^3 - c_1\right)^2}{9t^2\left(c_1+c_2t^3\right)^2}, \qquad c_1, c_2 \in \R.
\enqn
A straightforward computation shows that there indeed exists a phantom region for $t\in(t_1, t_2)$, where
\beqn
t_1=\sqrt[3]{\frac{4-3\sqrt{2}}{2}\cdot\frac{c_1}{c_2}},\qquad
t_2=\sqrt[3]{\frac{4+3\sqrt{2}}{2}\cdot\frac{c_1}{c_2}},
\enqn
and for simplicity we have assumed that $c_1 /c_2 >0$. Now picking two of these solutions (in our case -- two different solutions for the scale factor $a$) with the opposite values of constants $c_1$ and $c_2$ (say, choosing them to be negative for the first solution and positive for the second one), then performing an adequate time translation ($t \to t + t_0$, where $t_0$ is uniquely defined by the initial conditions) on the second of them, one would end up with two models, the first of which at $t=t_1$ delves into the phantom region, while the second one effectively pops out of it. All that is left for us to do then is to {\em sew} these two solutions together at $t=t_1$, thereby completely discarding the phantom regions altogether!

\begin{figure}
\begin{center}
\includegraphics[scale=0.7]{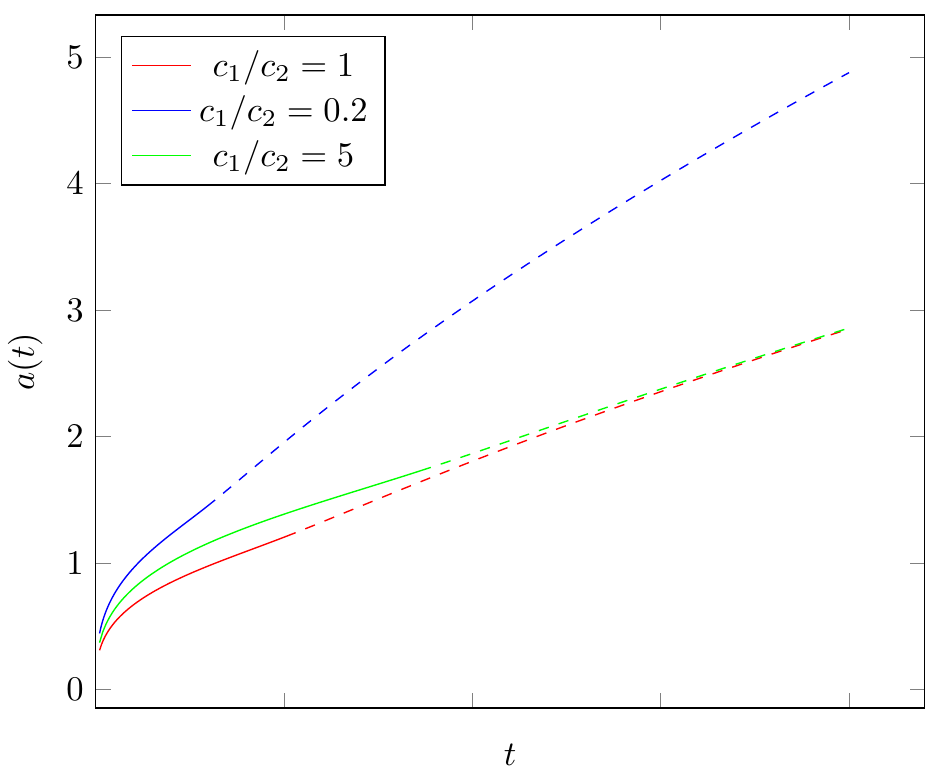}\includegraphics[scale=0.7]{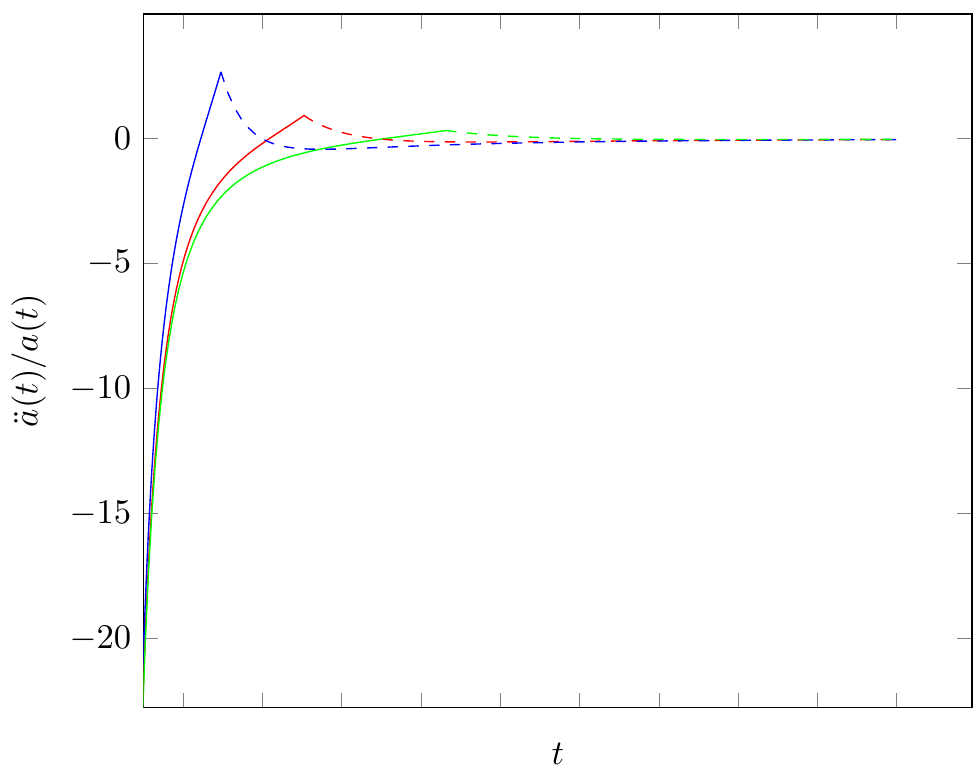}\\
\caption{The scale factor (left) and acceleration $\ddot{a}/a$ (right) as function of time for three values of $c_1/c_2$.} \label{Fig1}
\end{center}
\end{figure}

\begin{figure}
\begin{center}
\includegraphics[scale=0.7]{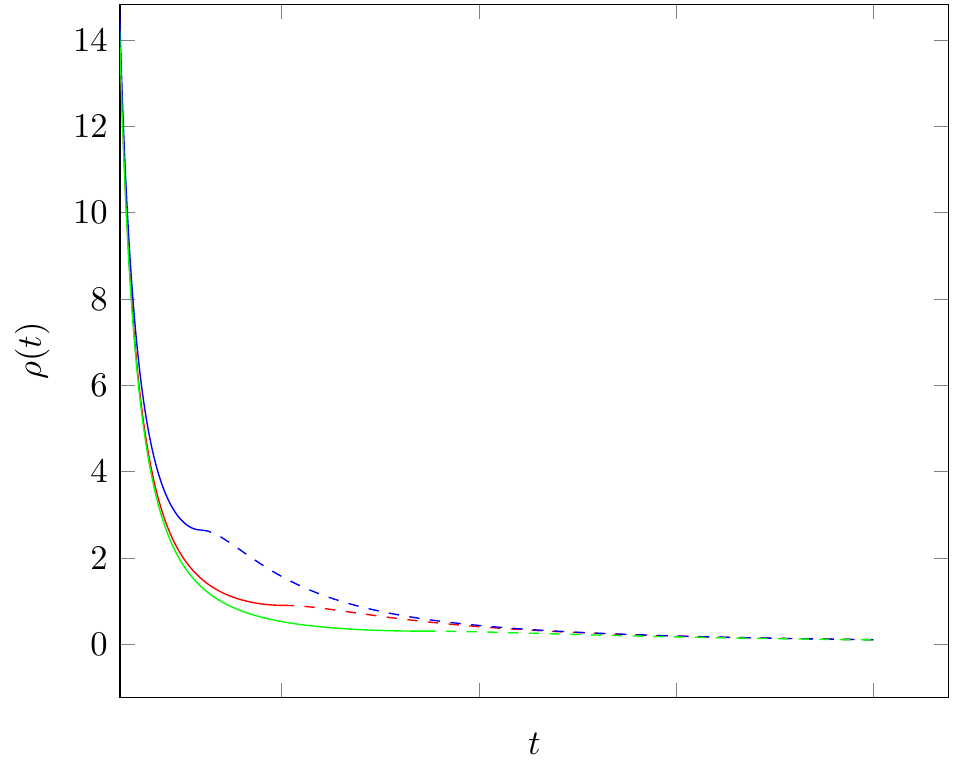}\includegraphics[scale=0.7]{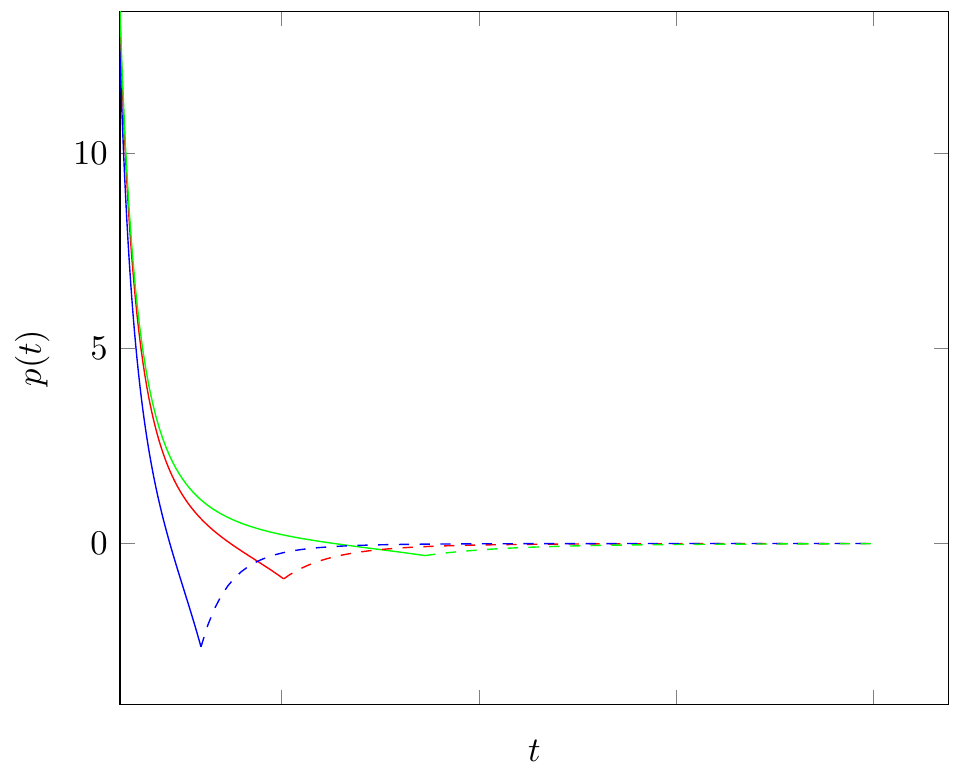}\\
\includegraphics[scale=0.7]{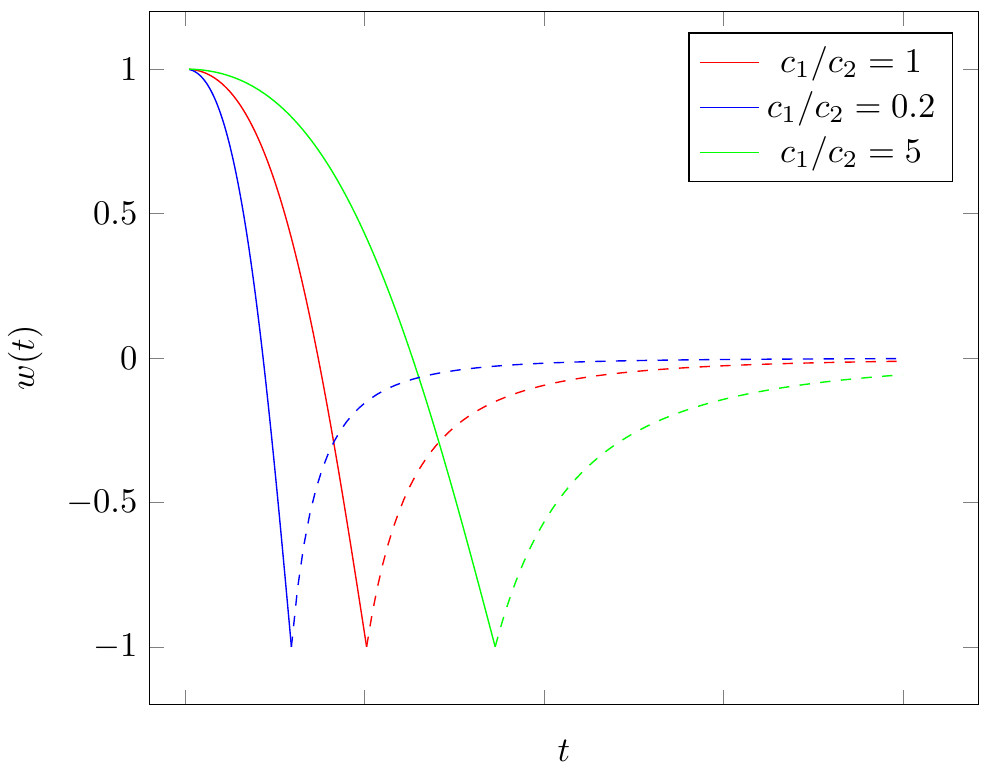}\\
\caption{Density $\rho(t)$, pressure $p(t)$ and the parameter $w=p/\rho$ as the functions of time for three values $c_1/c_2$.} \label{Fig2}
\end{center}
\end{figure}

The results of this operation for the scale factor $a$, acceleration parameter $\ddot a /a$, density $\rho$, pressure $p$ and the parameter of state $w$, performed for three particular values of $c_1/c_2$, are demonstrated on Figs. \ref{Fig1} and \ref{Fig2}.

One important thing we can immediately see here is the fact that, although $\ddot a/a$, $p$ and $w$ are all continuous at $t=t_1$, their derivatives are apparently not! In fact, the direct calculations show that no matter how one cuts and sews the functions $a_-$ and $a_+$ together, the third derivative of $a$, second derivative of $\rho$ and the first derivatives of $\rho$ and $w$ will {\em necessarily} have jump discontinuities at $t=t_1$. We believe that this result is actually very telling, but we will postpone the discussion of its importance for the corresponding Sec. \ref{sec:conclusion}. For now, let us just keep in mind that the jump points in the behaviour of the physical fields can be relevant as they are directly related to the problem of dephantomization.

With that said, let us now look at the phantom cosmologies in general (i.e. without any references to any scalar field) and see whether a little jump discontinuity might bring any sort of order there.
Let's consider for simplicity a universe filled with just dark energy. The equation of state for the dark energy can be written in the
following form:
\be
\label{EoS} p=-\rho-f(\rho)\, ,
\ee
where $f(\rho)$ is some function (we will get to the claims about its behaviour and continuity a bit later). For the phantom energy $f(\rho)>0$. Then
from the Friedmann equations, one can get the following expression for the time variable $t$:
\be \label{trho}
t-t_{0}=\frac{1}{3}\int^{\rho}_{\rho_0}\frac{d\rho}{\rho^{1/2}f(\rho)}.
\ee

Let's choose $t_{0}=0$ as a present time. If the universe expands, the phantom energy density starts increasing with time ($\rho>\rho_{0}$). It is
interesting to ponder the question of a possibility of dephantomization, i.e. of the equation of state crossing the line $w=-1$. There are two
possibilities.
\newline

{\bf Sudden dephantomization.}

Say, the function $f(\rho)$ has a jump discontinuity and changes its sign at the moment $t=t_{s}$. The second
derivative of the scale factor is undefined at the very moment of a jump. Such cosmological evolution can be called a {\em sudden dephantomization}
if $f(\rho)$ changes sign from ``+'' to ``--'' (similarly, it can be called a {\em sudden phantomization} if the change is reversed).

For illustration pick a solution for the phantom energy with a constant equation-of-state parameter $p=w_{0}\rho$,
$w_{0}=-1-\epsilon/3$, $\epsilon>0$:
\begin{equation}
a_{-}(t)=\frac{a_c}{(t_{BR}-t)^{2/\epsilon}},
\end{equation}
where $a_c$ is a constant and $t_{BR}$ is the time of a BRS (Big Rip singularity). But if we assume that at some moment $t_s$
a sudden dephantomization occurs, we would instead have a following solution for the scale factor
\begin{equation}
a_{+}(t)=a_s\exp(\lambda (t-t_s)),\quad t>t_s.
\end{equation}
One can choose parameters $a_c$, $\lambda$ and $t_s$ so that both the scale factor and its first derivative are continuous. The conditions are:
$$
\frac{a_c}{(t_{BR}-t_s)^{2/\epsilon}}=a_s,
$$
$$
\frac{2}{\epsilon(t_{BR}-t_s)}=\lambda.
$$
For given $a_c$ and $t_{BR}$ one can obtain $\lambda$ and $a_s$ as function of $t_s$. The second derivative of the scale factor has a
jump discontinuity at $t=t_s$:
$$
\frac{\ddot{a}_{+}(t_s)}{a_s}-\frac{\ddot{a}_{-}(t_s)}{a_s}=-\frac{2}{\epsilon(t_{BR}-t_s)^2}.
$$
The same process can also be performed for a universe filled with both the vacuum energy and the ordinary matter
after $t>t_s$. It would yield similar results.
\newline

{\bf Smooth dephantomization.}

Another interesting scenario to consider is the one of a {\em smooth dephantomization}. This scenario is possible when some particular value of density $\rho=\rho_{m}$ corresponds to $f(\rho_{m})=0$. In order for (\ref{trho}) to converge the function $f(\rho)$ in a close vicinity of the aforementioned point $\rho=\rho_{m}$ shall behave as $g(\rho)\sim C(\rho_{m}-\rho)^{\alpha}$, $0<\alpha<1$. Then, after reaching the separation line $w=-1$, two alternatives would exist: (i) universe
starts contracting and the energy density increases. Then the energy
density and the pressure (as a function of density) will be continuous at the moment $t=t_{s}$, but the first
derivative of the scale factor and the derivative $dp/d\rho$ would be ill defined. (ii) The universe continues to expand and the energy density
decreases. The equation of state essentially splits into two possible branches, with the dephantomization phenomena occurring exactly at the branch point.

Naturally, the scheme described above allows for an easy generalization. For the equations of state with the branch points the evolution of the universe undergoes the dephantomization and the so-called Quasi-Rip epochs.

For the purpose of illustration let's consider the following equation of state
\be \label{QR}
p=
\begin{cases}
-\rho-\frac{2}{3}\alpha^{2}(\rho_{m}-\rho)^{1/2}, & a\leq a_{T},\\
-\rho+\frac{2}{3}\alpha^{2}(\rho_{m}-\rho)^{1/2}, & a\geq a_{T},
\end{cases}
\ee
where $\rho\leq\rho_{m}$ and $a_{T}=a_{0}\exp(\rho_{m}^{1/2}/\alpha^{2})$ is a value of scale
factor at which the dephantomization occurs ($a_{0}$ is the scale factor at the moment when $\rho=0$). In this moment the value of
DE energy-density reaches the maximal value $\rho_{m}$.

For the scale factor as a function of time we have
$$
a(t)=a_{0}\exp\left(\frac{\rho_{m}^{1/2}}{\alpha^{2}}(1-\cos\alpha^2t)\right),\quad
0\leq t<\frac{\pi}{\alpha^2},
$$

At $t=\pi/2\alpha^{2}$ the dephantomization happens and the universe's expansion suffers a decceleration. Note, however, that at this moment (which corresponds to the universe crossing a phantom ``division line'') the derivative $dp/d\rho$ is ill defined. If we choose the EoS for the dark energy at $a>a_{T}$ in
another form it would be the third derivative of a scale factor at $t=t_{T}$ that ceases to exist. For example, pick
$$
p=-\rho,\quad t\geq t_{s}=\frac{\pi}{2\alpha^2}.
$$
After $t=t_{T}$ the universe expands according to de Sitter law
$$
a_{+}=a_{0}\exp\left(\frac{\rho_{m}^{1/2}}{\alpha^{2}}\right)\exp(\rho_{m}^{1/2}\left(t-t_{s}\right)),\quad
t\geq t_{T},
$$
so that a third derivative of a scale factor experiences a jump:
$$
\frac{d^{3}a_{+}(t_{s})}{dt^{3}}-\frac{d^{3}a_{-}(t_{s})}{dt^{3}}=
\alpha^{4}\rho_{m}^{1/2}a_{0}\exp\left(\frac{\rho_{m}^{1/2}}{\alpha^{2}}\right).
$$

\section{Conclusion and Discussion} \label{sec:conclusion}

In this article we have discussed a number of cosmological models predicting a whole new classes of singularities. The key factor in our derivations was the assumption  that either the scale factor or its derivatives might have a discontinuity of the first type. In particular, a new interesting class of singularities denoted as
``$w^c$-singularities'' arises when the scale factor is allowed to experience a finite jump at some moment of time. In the resulting model the first derivative of the scale factor diverges and so does the matter density -- and yet, despite all this, the parameter of state $w=p/\rho$ remains finite. Moreover, one can
actually construct the cosmological models with a finite pressure -- even at the threshold of singularity ($w=0$). For a scalar field the latter corresponds to a
finite Lagrangian density at the moment of singularity.

The possibility of a finite jump of the first derivative of a scale factor allows one to construct the cosmological models with what we have termed a ``survivable
sudden future'' singularity. If the model in question describes a ``bouncing'' universe in which the first derivative of a scale factor changes sign, the energy density is shown to remain continuous.

The consideration of the jump points for the higher order derivatives of a scale factor should not be overlooked, since they also leads to a number of interesting features. The jump suffered by the second derivative might correspond to a sudden dephantomization, i.e. the moment when a parameter of state suddenly
becomes equal to or exceed the value $w=-1$ (or vice versa if it is the phantomization we are after). As illustration we have considered a particular cosmological
evolution that instantaneously transits from the accelerated expansion phase governed by the phantom energy to the phase of a de Sitter expansion. The only possible downside would be the fact that the pressure in such a universe shall experience a finite jump at the very moment of aforementioned singularity. This, however, appears to be a rather small price to pay, compared to the fact that such a scenario actually allows for the evolution of the universe (not to mention its possible inhabitants) to be extended well beyond this singularity. The importance of this is further emphasized if one is to recall that the general solution of the cosmological equations for at least some scalar field potentials cross the phantom line only to experience the ultimate wrath of the phantom fields: the big rip singularity (BRS). But, as we have said before, it is possible to merge two such solutions with $w>-1$ completely cutting out both the phantom zone and the BRS. In this case it is the third derivative of the scale factor that will suffer a jump at the cut.

And so, we come to the last point of this article. As our reader no doubt recalls, we have been resolute in answering the first of the objections, summarized in Sec. \ref{sec:Intro}, namely: whether the inclusion of the jump points into the cosmological considerations yields any noticeable difference to the familiar cosmological scenarios. We entertain the hope that this article has provided well enough material to thoroughly refute any such doubts. However, it is the second objection we haven't really touched: whether these new phenomena, summoned by our introduction of the jump points, have any {\em plausibility} for the description of the real material universe. After all, is it not possible that the models henceforth described are but a mathematical trifle, a curiosity with no physical meaning whatsoever?.. We do not think so, and we base our reasoning on two observations. First is the phenomena of the {\em crossing to the phantom zone}. Second one deals with recent works that cast new doubts on the validity of the Strong Cosmic Censorship principle (SCC).

Let us start with the former. We already know that many of the physically plausible scalar field potentials produce the cosmological dynamics that at some moment of time pushes the universe into the phantom zone, thereby changing the sign of the kinetic term $\dot \phi^2/2$. If we are to accept this, then we are also forced to conclude that the function $\dot \phi$ actually becomes {\em imaginary} -- a conclusion that would be a very hard pill to swallow for any physicist. It is possible to avoid this, but only by means of literal cutting-and-pasting of two cosmological solutions, thus eliminating both the phantom zone and its unpleasant constituents (including, but not limited to the Big Rip Singularity). However, as we have just discussed, such a mathematical operation leaves a mark on the behaviour of the scale factor: its third order derivative must experience a jump!.. Therefore, we are facing two possibilities: either to accept the existence of the phantom zone and treat the negativity of the kinetic term as physical (which is frankly a rather dim perspective), or to reject it as unphysical by accepting the possibility of a single jump discontinuity in at least some of the cosmological variables (scale factor, pressure, density and/or their derivatives). We believe the latter option to be much more palatable, but we leave the final decision to our reader.

Now let us turn to the second observation, which revolves around the deterministic character of General Relativity Theory (GR). It is a known fact that the ultimate plague of any deterministic theory is the growth of the Cauchy horizons in its models. In particular, an emergence of a Cauchy horizon in the solutions of the Einstein equations poses a serious threat for the deterministic nature of GR, since the dynamics of any object (observer) after it crossed the Cauchy horizon can not be predicted, even in theory. Until recently the usual repartee to this problem was that such a ``pathologic'' behaviour shall be endemic solely to the equally ``pathologic'' regions of space-time -- the black holes, where it will be essentially nullified by the Penrose's strong cosmic censorship (SCC) principle. However, one recent article \cite{CCDHJ} has cast a serious doubt on this reasoning. The article in question was dedicated to studying the massless scalar fields in the exterior of the Reissner-Nordstr\"om-de Sitter black holes (RNdS) by means of the computation of the quasinormal modes (QNMs). The conclusion was nothing short of staggering: it appears that SCC might actually be violated! If this conclusion will hold in the subsequent inquiries, this would mean that inside of RNdS black holes the General Relativity ceases to be deterministic.

This rather unexpected discovery will undoubtedly have a lot of ramifications for various areas of physics, but here we are interested in just one: the effect the SCC violation might have on the global cosmology at large. In order to understand this effect (and to figure out why there might be any at all) let us consider a Hubble volume of radius $H^{-1}$ situated in a flat Friedmann universe. Say, we want to calculate the total mass of this volume. Naturally, it will be equal to the product of the total volume, which is proportional to $H^{-3}$ and the critical density, proportional to $H^2$. Then we ask ourselves: what would be the gravitational radius of this Hubble volume? Amazingly, the answer is {\em exactly} $H^{-1}$! This surprising fact led different researches to reason that the argumentation originally developed for the black holes might actually be applicable to the observable universe; one might, for example, study the holographic dynamics of our universe or even derive the Friedmann equations from the holographic principle. Furthermore, it is conceivable that similar arguments might relate the observable universe with such particular species of black holes as RNdS black holes, or, alternatively, that the mechanisms that allow for an SCC violations would also be discovered in other types of black holes (perhaps those better ``suited'' to reflect the parameters of our observable universe).~\footnote{Of course, it has to be owned that we are still quite far from understanding the intriguing correspondence between the gravitational radius of our universe and the horizon's radius $H^{-1}$. Until we actually reach that point, the claims about the ``similarities'' between our universe and the black holes of certain types shall only be acknowledged in a strictly Pickwickian sense \cite{Dickens}.} If that would be the case, any proper cosmological model will have to include a certain point $t_s$ where a break in purely deterministic evolution will occur, manifesting itself as a jump in either the scale factor of some of its higher derivatives at the very moment $t=t_s$.

Indeed, if we are forbidden from analytically continuing the solution beyond the moment $t_s$, then in order to actually describe the dynamics of the universe at that moment we'll have to to construct it by joining the solution $a_{-}(t)$ (for $t<t_s$) at $t=t_s$ with the solution $a_{+}(t)$, chosen from the (probably infinite) class of solutions for $t>t_s$. This very act then would naturally produce a solution whose $n$-th order derivative $d^n a/dt^n$ will have a jump discontinuity at $t=t_s$. The easiest way to visualise this would be via the already discussed smooth dephantomization from Sec. \ref{sec:dephantom}. Indeed, consider an exact solution of the Friedmann equations which describes two consequent phantom zones. The crossing of the phantom zone I occurs at $t^{(I)}_1$ (the beginning of the phantom zone) and $t^{(I)}_2$ (when the universe ceases to be phantom), whereas the phantom zone II will stretch from $t=t^{(II)}_1$ to $t=t^{(II)}_2$ (naturally, $t^{(I)}_1<t^{(I)}_2<t^{(II)}<t^{(II)}_2$). Both of those phantom zones might produce their own Big Rip singularities, one belonging to the interval $t \in \left(t^{(I)}_1; t^{(I)}_2\right)$, and another at $\left(t^{(II)}_1; t^{(II)}_2\right)$. We can eradicate both by eliminating the corresponding phantom zones via the formalism of dephantomization from Sec. \ref{sec:dephantom}. However, there is not one, but {\em two} different ways to perform such an operation. On the one hand, we can surgically remove two phantom zones separately from each other, first suturing the points $t^{(I)}_1$ with $t^{(I)}_2$, and then $t^{(II)}_1$ with $t^{(II)}_2$. Then again, we can perform one major operation, stitching together the points $t^{(I)}_1$ and $t^{(II)}_2$. What kind of solutions will we end up with? In the first case it will be a solution with two jump discontinuities of the third order derivative of $a(t)$, and the second solution will have just one. By uniting them together with the Big Rips-infested original, we therefore end up with not one or two, but {\em three} different cosmological solutions stemming from the {\em identical initial conditions} and, in fact, sharing exactly he same history up to the point $t=t^{(I)}_1$, yet seriously diverging there. This perfectly exemplifies the types of dynamics one can expect at the moments of a breakage of determinacy in the scale factor and serves as a fitting point to conclude our discussion.

\acknowledgements

The authors would like to express their gratitude to professor John D. Barrow for his interest in the article and for his helpful suggestions. The work is supported by project 1.4539.2017/8.9 (MES, Russia).

\end{document}